\documentclass[twocolumn,showpacs,preprintnumbers,amsmath,amssymb]{revtex4}
\usepackage{graphicx}
\usepackage{dcolumn}
\usepackage{bm}

\newcommand{\bef}{\begin{figure}}
\newcommand{\eef}{\end{figure}}

\newcommand{\be}{\begin{equation}}
\newcommand{\ee}{\end{equation}}
\newcommand{\bea}{\begin{eqnarray}}
\newcommand{\eea}{\end{eqnarray}}
\begin{document}

\title{Energy dependence of $\bar{p}/p$ ratio in $p$+$p$ collisions}

\author{Subhash Singha$^1$, Pawan Kumar Netrakanti$^2$, Lokesh Kumar$^3$ and Bedangadas Mohanty$^1$ }
\affiliation{$^1$Variable Energy Cyclotron Centre, Kolkata 700064, India, \\
$^2$Bhabha Atomic Research Centre, Mumbai 400 085, India and  \\ 
$^3$Kent State University, Kent, Ohio 44242, USA}
\date{\today}
\begin{abstract}
We have compiled the experimentally measured $\bar{p}/p$ ratio 
at midrapidity in $p$+$p$ collisions from $\sqrt{s}$ = 23 to 7000 GeV 
and compared it to various mechanisms of baryon production as implemented in  
PYTHIA, PHOJET and HIJING/B-$\bar{\mathrm B}$ models. For the models studied with 
default settings, PHOJET has the best agreement with the measurements, 
PYTHIA gives a higher value for $\sqrt{s}$ $<$ 200 GeV and the ratios from
HIJING/B-$\bar{\mathrm B}$ are consistently lower for all the $\sqrt{s}$ studied. 
Comparison of the data to different mechanisms of
baryon production as implemented in PYTHIA shows that through a suitable 
tuning of the suppression of diquark-antidiquark pair production
in the color field relative to quark-antiquark production
and allowing the diquarks to split according to the popcorn scheme 
gives a fairly reasonable description of the measured $\bar{p}/p$ ratio for
$\sqrt{s}$ $<$ 200 GeV. Comparison of the beam energy dependence of the 
$\bar{p}/p$ ratio in $p$+$p$ and nucleus-nucleus (A+A) collisions at 
midrapidity shows that the baryon production is significantly more 
for A+A collisions relative to $p$+$p$ collisions for  
$\sqrt{s}$ $<$ 200 GeV. We also carry out a phenomenological fit to 
the $y_{\mathrm {beam}}$ dependence of the $\bar{p}/p$ ratio.
\end{abstract}
\maketitle

Protons ($p$) and anti-protons ($\bar{p}$) are the most abundantly produced
baryons in high energy collisions. These have been measured at various
center of mass energies ($\sqrt{s}$) in hadron-hadron~\cite{ISR,alice} and nucleus-nucleus 
collisions~\cite{AA} as a function of rapidity ($y$) and transverse momentum 
($p_{\mathrm T}$). Rapidity dependence of baryon production is expected to provide information 
on baryon transport and stopping~\cite{brahms} and the $p_{\mathrm T}$ 
dependence of the yields is expected to help in understanding the baryon 
production mechanism~\cite{alice}. 
The $\bar{p}/p$ ratio within the assumption of a thermal model is used 
to obtain the baryon chemical potential in heavy-ion collisions~\cite{cleymans,star}. 
Recently it has been argued based on QCD that there are constraints on
allowing quarks to trace the baryon number~\cite{kharzeev}, although they carry 
a baryon number of 1/3 based on quark model classification.
It is argued that the trace of the baryon number could be associated with 
non-perturbative configurations of gluon fields rather than to valence quarks. 
All these make the study of the energy dependence of $\bar{p}/p$ ratio important 
in high energy collisions, where the role of the gluonic contributions to particle 
production is expected to increase with $\sqrt{s}$. 

In the string picture the process of baryon production is not unique.
Mesons in such a picture can be viewed with a short string between 
a quark and anti-quark endpoints. However for the baryons consisting of 
three quarks it is difficult to visualize in a simple way. 
Baryon production in string
picture is implemented in the PYTHIA model~\cite{pythia}. The simplest 
mechanism of baryon production in such a picture is through a diquark model.
Any quark of a given flavor is assumed to be represented either by a quark 
or an antidiquark in a color triplet state. Then the baryon and 
antibaryon are produced as nearest neighbors along the string. Such a model
has to deal with the relative probability to pick a diquark 
over a quark. The extra suppression associated with a diquark containing 
a strange quark purely from phase space considerations and when a baryon 
is formed by joining a diquark and a quark, it has to be a symmetric 
three-quark state. Another equivalent mechanism is that in which diquarks 
as such are never produced, but baryons appear due to the successive production 
of several quark-antiquark pairs. Such a mechanism is referred to as the {\it popcorn}
mechanism. These pairs exist by means of the 
color fluctuations in the field~\cite{popcorn}. An advanced version of the 
popcorn mechanism is described in Ref.~\cite{apopcorn}. While the simpler popcorn  
mechanism admits at most one intermediate meson formation, the advanced version, 
on the other hand, allows for the possibility of many such mesons.

Another model of particle production which is widely used for comparison 
to data is the PHOJET~\cite{phojet}. The PHOJET, is a two component model 
that combines the ideas of the Dual Parton Model (DPM)~\cite{dpm}
(soft processes) with perturbative QCD (hard processes). 
The mechanism of Pomeron exchange is at the heart of the DPM. 
According to the DPM, the leading contribution to multiparticle production in high-energy 
hadron-hadron collisions arises from the exchange of a single Pomeron 
between the colliding hadrons. Secondary Pomeron exchanges account for the remaining 
activity in the event. Each exchanged Pomeron gives rise to two color-neutral 
chains stretching between quarks and diquarks, for baryons, 
or quarks and anti-quarks for mesons.
The basic 
difference between PYTHIA and PHOJET lies in their approach towards an event
formation. The starting point of particle production in PYTHIA is through the 
description of possible hard interactions in $e^{+}+e^{-}$, $p$+$p(\bar{p})$ or $e$+$p$ colliders 
and then combines several ideas for the soft hadronic interactions, whereas in PHOJET model it 
initializes the event generation by describing the soft component of 
hadron-hadron, photon-hadron or photon-photon interactions at high energies. 
The hard component is introduced later and calculated by perturbative QCD at the partonic level.

A novel mechanism of baryon transport motivated by the Regge theory~\cite{regge} 
and differing from the diquark breaking model has been implemented in the 
form of an event generator, the HIJING/B-$\bar{\mathrm B}$~\cite{hijingbb}. 
As discussed briefly earlier, the mechanism is motivated from 
the non-perturbative gluon field configuration called the the baryon junction.
The baryon junction is found to be originating from the basic concepts of
QCD and is a vertex where the color flux lines flowing from the 
three valence quarks are connected. The junction is expected to play a dynamical role 
through the Regge exchange of junction states in high
energy collisions. The junction exchange could  provide 
a natural mechanism for the transport of baryon number into the central rapidity region.
Further details can be found in Ref.~\cite{kharzeev, hijingbb}

In this article we compare the energy dependence of the experimentally measured 
$\bar{p}/p$ ratio at midrapidity in $p$+$p$ collisions for various $\sqrt{s}$ to the above
discussed models. The data for the $\sqrt{s}$ = 23, 31, 45,
53 and 63 GeV are from the ISR experiments~\cite{ISR}. The $\bar{p}/p$ ratio at 
$\sqrt{s}$ = 200 GeV is from the STAR experiment at RHIC~\cite{star}. And the 
$\bar{p}/p$ ratio at the highest energies of 900 GeV and 7 TeV are from
ALICE at LHC~\cite{alice}. 

\bef
\begin{center}
\includegraphics[scale=0.4]{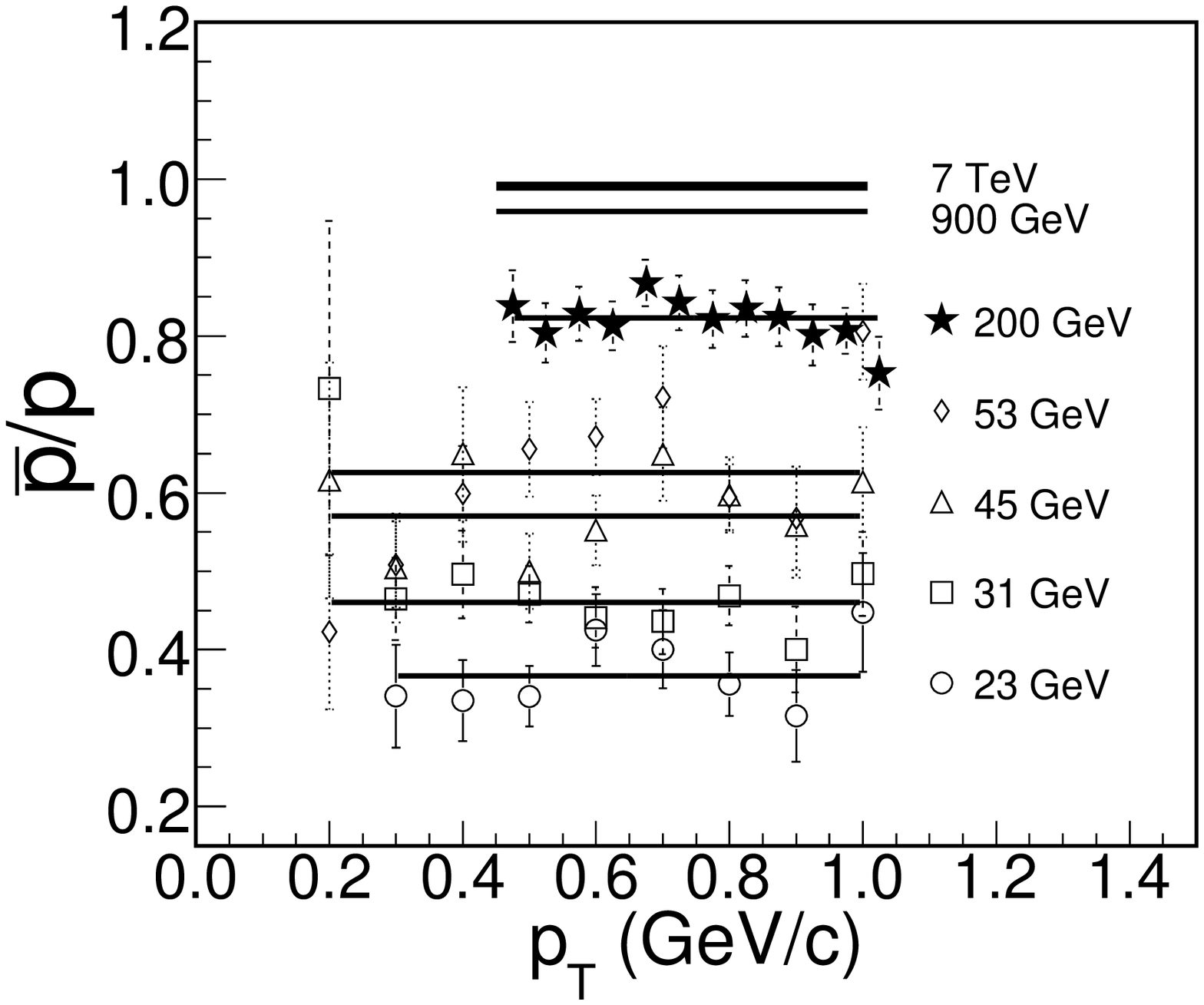}
\caption{$\bar{p}/p$ ratio at midrapidity as a function 
of $p_{\mathrm T}$ for $p$+$p$ collisions. The solid lines
corresponds to new data from ALICE at LHC~\cite{alice}. The
$\bar{p}/p$ ratio for $\sqrt{s}$ = 200 GeV is from the 
STAR experiment at RHIC~\cite{star}. The data for  $\sqrt{s}$ $<$ 200 GeV
are from the ISR~\cite{ISR}. The
dashed lines are straight line fits to the $p_{\mathrm T}$ 
dependence of the $\bar{p}/p$ ratios at various $\sqrt{s}$,
assuming the ratios do not depend on $p_{\mathrm T}$ for the measured range.}
\label{fig1}
\end{center}
\eef
Figure~\ref{fig1} shows the $\bar{p}/p$ ratio at midrapidity
for various $\sqrt{s}$ (except $\sqrt{s}$ = 63 GeV, reasons for not showing 
these are discussed later),
in $p$+$p$ collisions as a function of $p_{\mathrm T}$. All data points are
only shown for the $p_{\mathrm T}$ $<$ 1 GeV/$c$. We observe that the ratios
are constant as a function of $p_{\mathrm T}$ for each $\sqrt{s}$  and 
the value of the $\bar{p}/p$ ratio increases with  $\sqrt{s}$. 
The solid lines show the recent measurements by the ALICE experiment at LHC~\cite{alice}. 
The low $p_{\mathrm T}$ dependence of
the $\bar{p}/p$ ratio is discussed in Ref.~\cite{alice}, where a drop of 
$\bar{p}/p$ with increase in $p_{\mathrm T}$ is expected in HIJING/B-$\bar{\mathrm B}$.
In the this paper we discuss only the energy dependence of the $\bar{p}/p$ ratio 
at midrapidity by comparison to various models of baryon production.

 We first compare the experimental measurements to 
PYTHIA, PHOJET and HIJING/B-$\bar{\mathrm B}$ models, then compare the $\bar{p}/p$ ratio
to various baryon production mechanism as implemented in PYTHIA. This is followed
by the discussion of asymmetry in proton and anti-proton production and comparison
to corresponding available results from heavy-ion collisions.

\bef
\begin{center}
\includegraphics[scale=0.4]{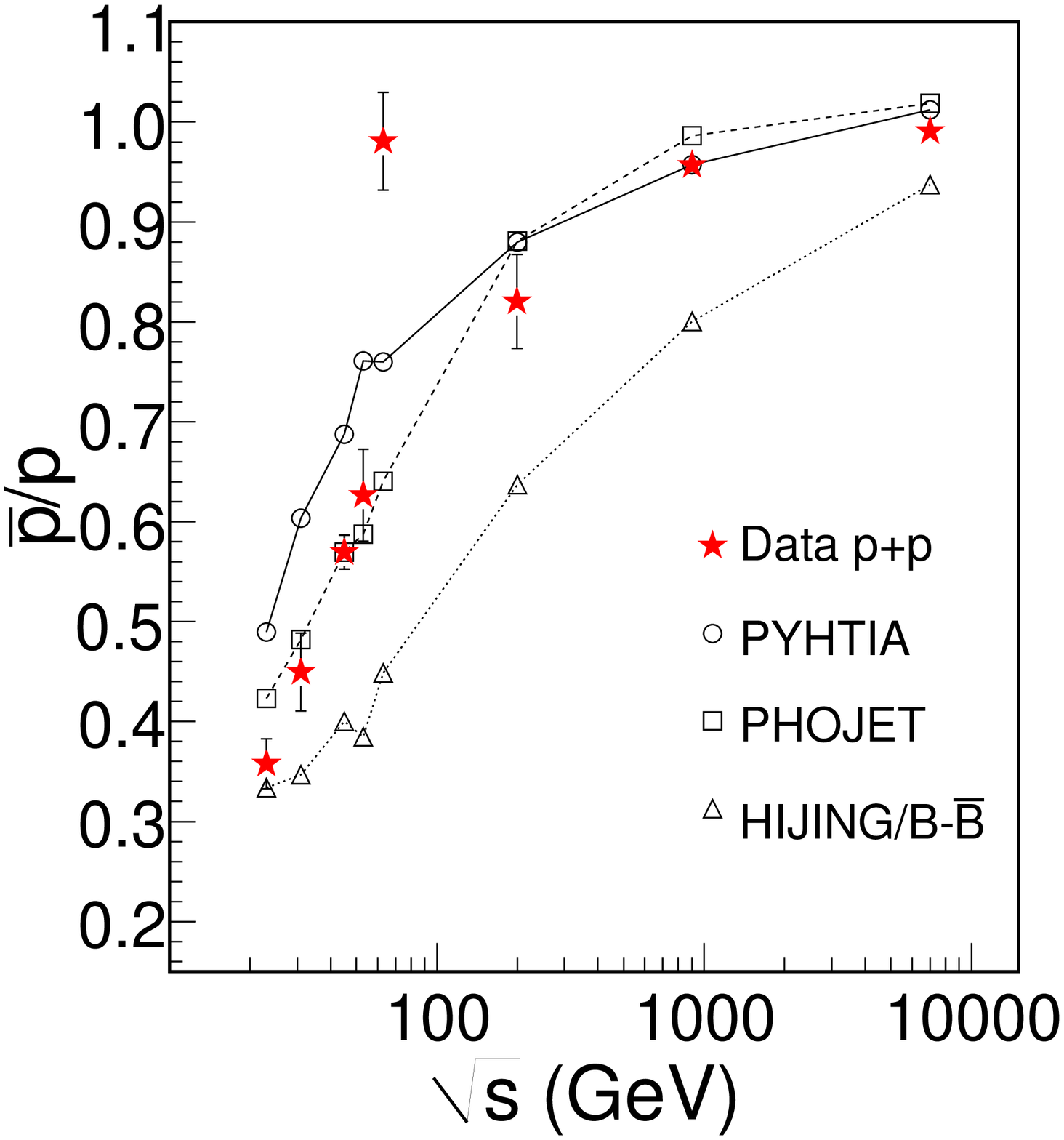}
\caption{(Color online) $\bar{p}/p$ ratio at midrapidity as a function 
of $\sqrt{s}$ for $p$+$p$ collisions. The experimental 
data are compared to model calculations 
from PYTHIA~\cite{pythia}, PHOJET~\cite{phojet} 
and HIJING/B-$\bar{\mathrm B}$~\cite{hijingbb}
with default settings.}
\label{fig2}
\end{center}
\eef
Figure~\ref{fig2} shows the increase of the $p_{\mathrm T}$ integrated 
$\bar{p}/p$ ratio at midrapidity 
with increase in $\sqrt{s}$ for $p$+$p$ collisions. The experimental
data are compared to three models,viz, PYTHIA (Ver. 6.4),
PHOJET (Ver. 1.12) and HIJING/B-$\bar{\mathrm B}$ (Ver. 1.34),all with default settings. The $\bar{p}/p$ ratio for
$\sqrt{s}$ = 63 GeV shows an abnormally high value, although shown in the figure,
we would not consider it for physics discussions. It is expected that RHIC data
collected in the year 2005 at $\sqrt{s}$ = 63 GeV will help in resolving the abnormality in the $\bar{p}/p$ ratio. All models studied show that the $\bar{p}/p$ ratio
increases with  $\sqrt{s}$ and approaches unity for higher energies (LHC). 
Infact, the PYTHIA and PHOJET models give very similar values
at the LHC energies, while HIJING/B-$\bar{\mathrm B}$ under predicts the
$\bar{p}/p$ ratio. 
The major difference occurs for $\sqrt{s}$ $<$ 200 GeV, PYTHIA model
gives higher values and HIJING/B-$\bar{\mathrm B}$ continues to give lower values of 
 $\bar{p}/p$ ratio. Only the PHOJET model with default settings gives a 
reasonable description of the $\sqrt{s}$ dependence of the measured 
$\bar{p}/p$ ratio for $p$+$p$ collisions.

\bef
\begin{center}
\includegraphics[scale=0.4]{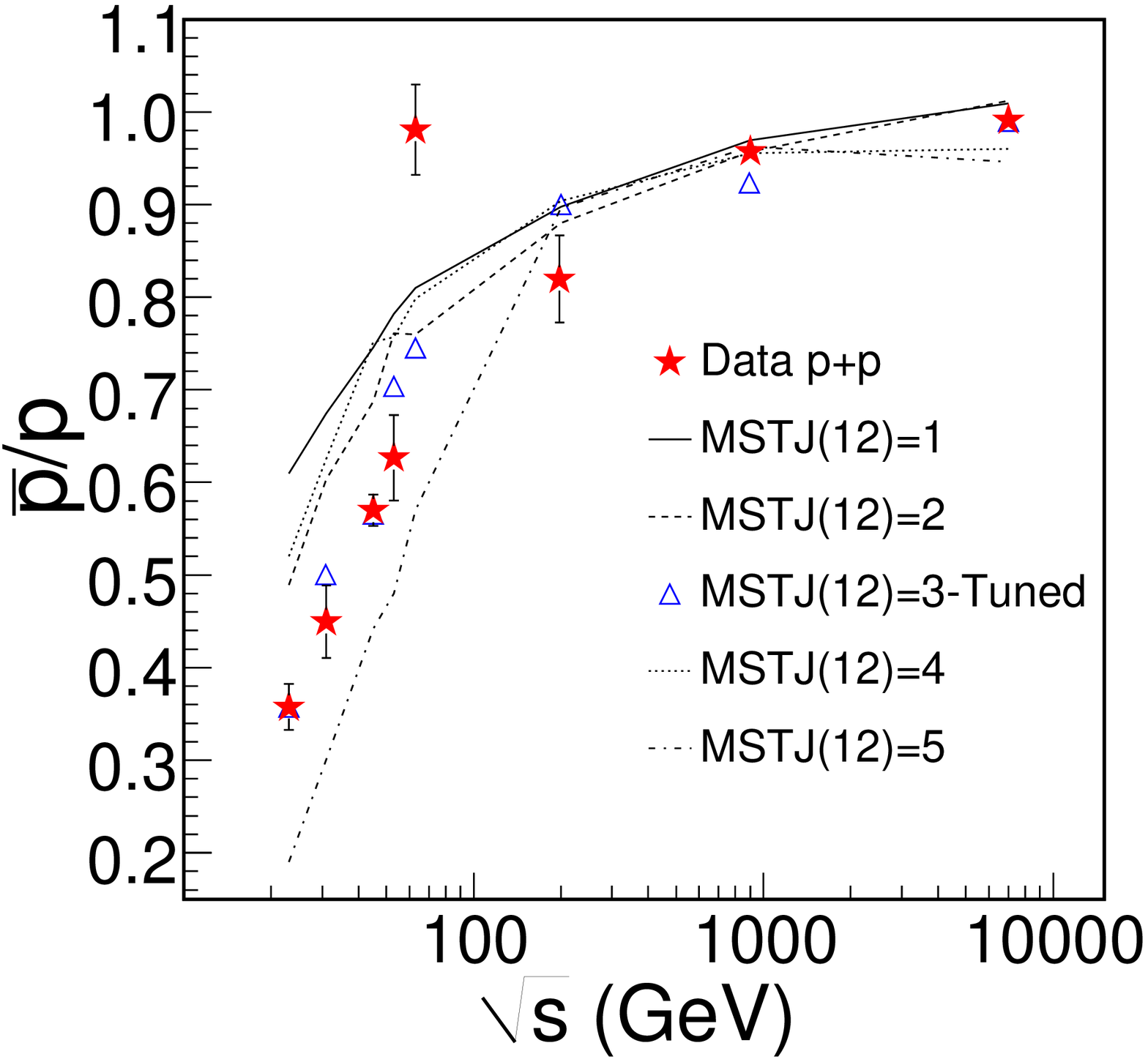}
\caption{(Color online) $\bar{p}/p$ ratio at midrapidity as a function 
of $\sqrt{s}$ for $p$+$p$ collisions compared to various
implementation of the baryon production schemes in PYTHIA. See
text for more details.}
\label{fig3}
\end{center}
\eef

The PYTHIA model has some variations in the baryon production
mechanism. In Fig.~\ref{fig3} we compare the 
experimental $\bar{p}/p$ ratio to such variations as 
implemented in the model. The parameter that we varied
is known as MSTJ(12), it can take up values from 0 to 5,
with the value of 2 as the default setting (one used in
Fig~\ref{fig2}). We did not consider MSTJ(12) = 0, as it
corresponds to no baryon-antibaryon pair production.
The condition MSTJ(12) = 1 refers to the mechanism where
baryon production is through diquark-antidiquark pair production with
the diquark being treated as a unit. While MSTJ(12) = 2 has the additional 
possibility for diquark to be split according to the popcorn scheme.
The mechanism of baryon production for the case MSTJ(12) = 3 is 
same as that for MSTJ(12) = 2, but has an  additional condition that 
the production of first rank baryons may be suppressed. For this
case, we additionally changed the value of the parameter which
governs the suppression of diquark-antidiquark pair production
in the color field, compared with quark-antiquark production.
The value we put in for this parameter is 0.05 compared to the
default value of 0.1. This is referred to as MSTJ(12)=3-Tuned in
the Fig.~\ref{fig3}. The condition MSTJ(12) = 4 again revolves 
around MSTJ(12) = 2 with an extra condition that the diquark vertices 
are suppressed. The last scheme implemented corresponds to 
MSTJ(12)= 5 is similar to MSTJ(12) = 2, but with an advanced version of 
the popcorn model. Fig.~\ref{fig3} shows that for the lower beam
energies the model results with the condition MSTJ(12)=3-Tuned has a 
reasonable agreement with the experimental data. For higher 
$\sqrt{s}$ all the above conditions give similar values of $\bar{p}/p$ ratio.

\bef
\begin{center}
\includegraphics[scale=0.4]{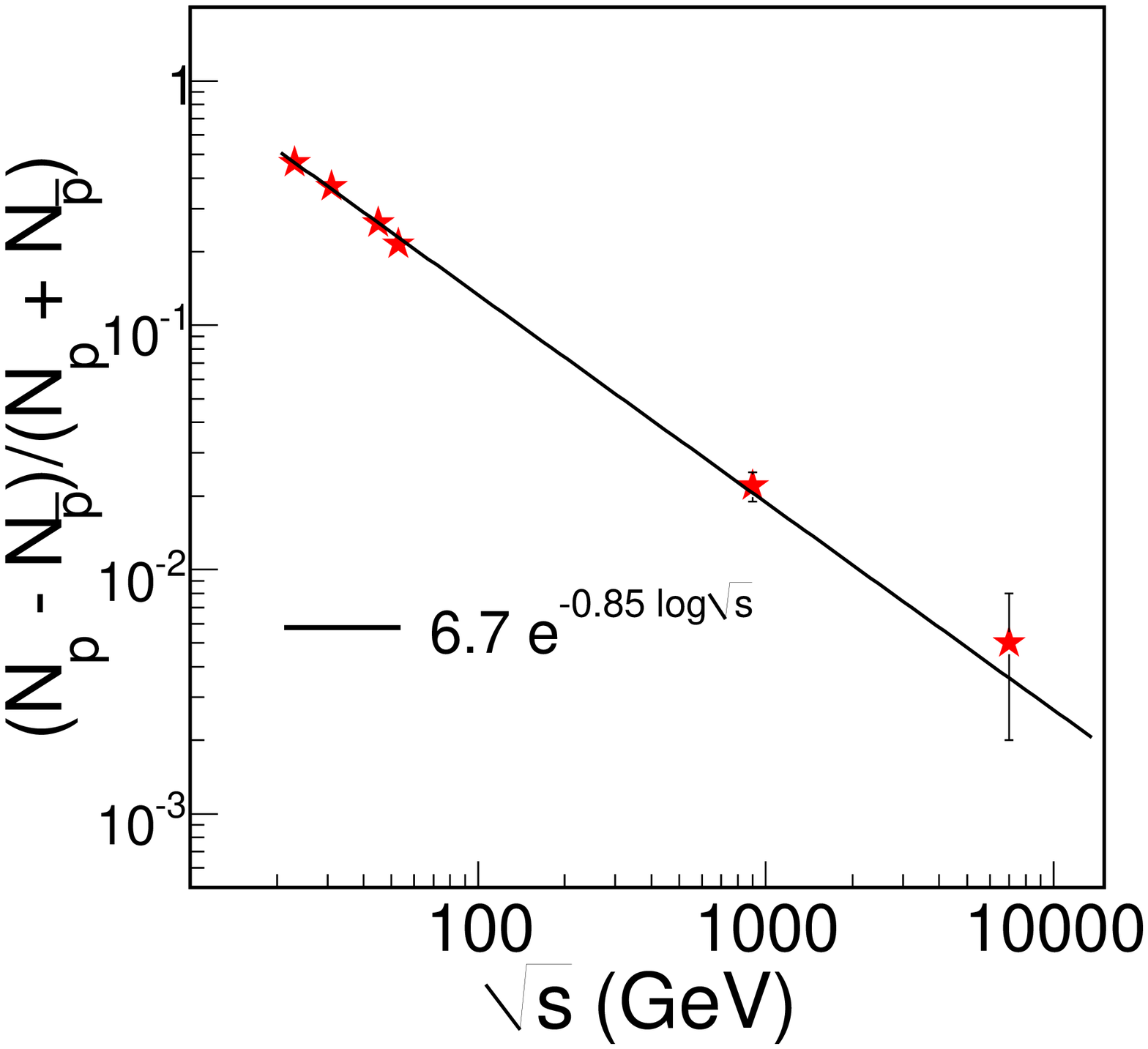}
\caption{(Color online) Asymmetry for proton and anti-proton production at 
midrapidity for $p$+$p$ collisions as a function of $\sqrt{s}$.
The solid line is a fit to the data, with the functional form shown.}
\label{fig4}
\end{center}
\eef

Different baryon production mechanisms could lead to an asymmetry 
in the production of protons and anti-protons. This asymmetry
can be measured by constructing the following ratio,
\begin{equation}
\frac{N_{p} - N_{\bar{p}}}{N_{p} + N_{\bar{p}}},
\end{equation}
where $N_{p}$ and $N_{\bar{p}}$ are the number of protons and
anti-protons. As pair production would lead to same number of protons 
and anti-protons, the asymmetry will have a value of zero. Any
non-zero value indicates the fraction of protons in midrapidity due
to effects such as stopping. Figure~\ref{fig4} shows the asymmetry
ratio for protons and anti-protons as measured in $p$+$p$ collisions
for various $\sqrt{s}$ ranging from 23 GeV to 7 TeV. The asymmetry
is found to decrease with increase in $\sqrt{s}$, indicating 
the decreasing contributions of protons due to stopping at midrapidity.
The ratio changes from about 46\% at $\sqrt{s}$ = 23 GeV to 0.5\% at
the top LHC energy of 7 TeV. This range in $\sqrt{s}$ corresponds to
a range in $y_{\mathrm {beam}}$ of 3 to 9 units, respectively.  This information 
is useful to study double baryon production in $p$+$p$ collisions~\cite{kharzeev}
and baryon number flow over long rapidity interval~\cite{baryonflow}.
The solid line in Fig.~\ref{fig4} is a fit ($\chi^{2}/ndf$ = 3/4) 
to the experimental data with the function $Ae^{-Blog \sqrt{s}}$, 
with the parameters $A$ = 6.7 $\pm$ 0.9 and $B$ = 0.85 $\pm$ 0.04.

\bef
\begin{center}
\includegraphics[scale=0.4]{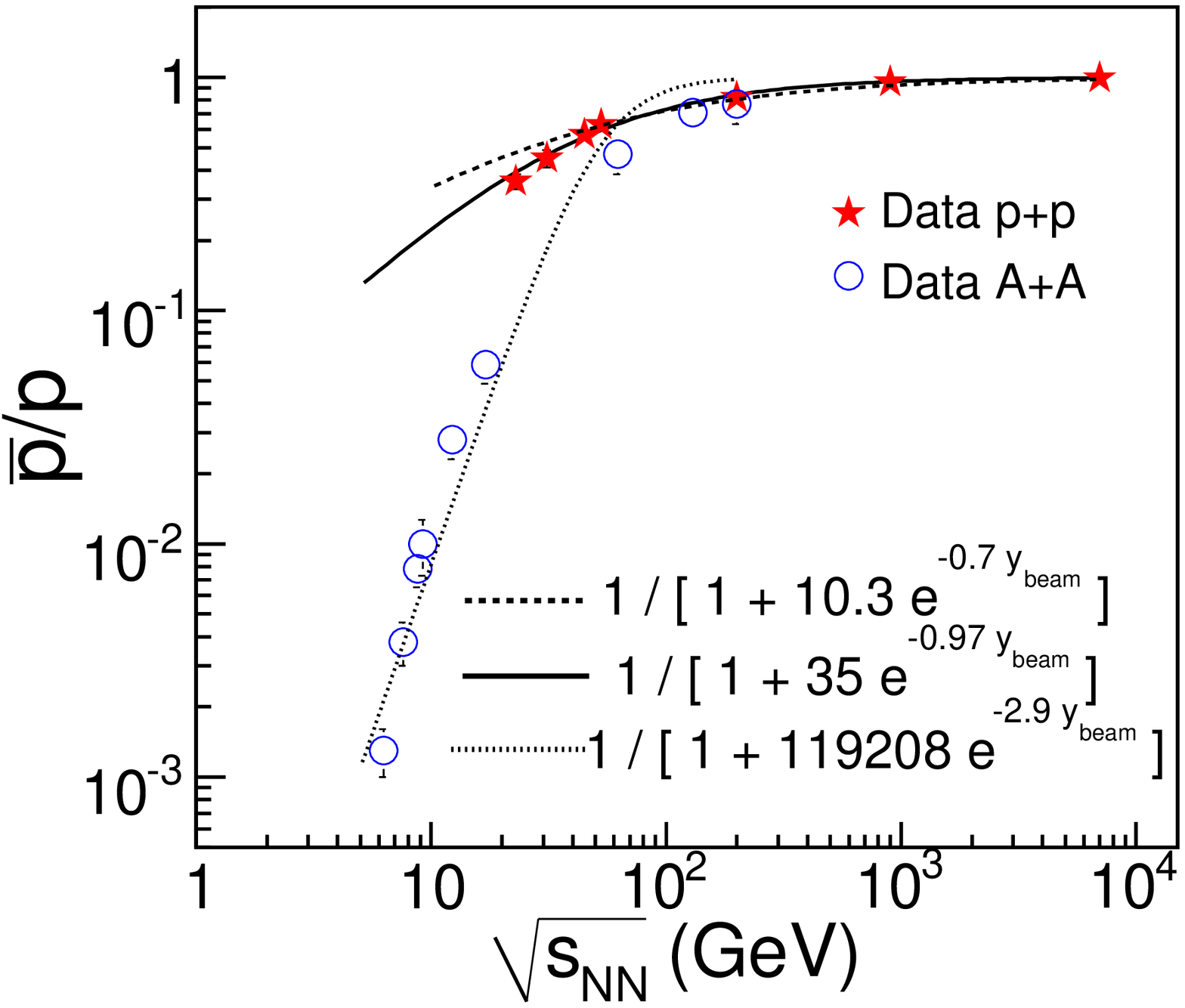}
\caption{(Color online) Comparison of beam energy of $\bar{p}/p$ ratio at midrapidity 
for $p$+$p$ and nucleus-nucleus (A+A) collisions. The solid, dash and dotted 
lines are fits to the $p$+$p$ and A+A data inspired by the model 
based on baryon string junction picture~\cite{kharzeev,kaidalov}.}
\label{fig5}
\end{center}
\eef
In Fig.~\ref{fig5} we compare the beam energy dependence of the experimentally 
measured midrapidity $\bar{p}/p$ ratio in $p$+$p$ collisions to the available
$\bar{p}/p$ ratios at midrapidity in nucleus-nucleus (A+A) collisions. The 
A+A collision data are taken from the experiments at AGS~\cite{ags}, SPS~\cite{sps} 
and RHIC~\cite{AA}. For both the systems the $\bar{p}/p$ rapidly rises with 
beam energy and approaches unity. For the $p$+$p$ collisions the $\bar{p}/p$ ratio
has a value of 0.991 $\pm$ 0.005(stat.) $\pm$ 0.014(syst.) at 7 TeV, 
while for heavy-ion collisions the  $\bar{p}/p$ ratio
has a value of 0.77 $\pm$ 0.14 at 200 GeV. 
Looking at the region in beam energy where there is overlap between
$p$+$p$ and A+A collisions, the relative proton contributions at midrapidity for 
A+A collisions is more than for $p$+$p$ collisions. The values of the ratio seem
to become equal around 200 GeV. Also shown in the figure are fits 
to the experimental $\bar{p}/p$ ratio in the $p$+$p$ collisions 
to a function of the form 
$ [1 + C exp[(\alpha_{J} - \alpha_{P})y_{\mathrm {beam}}]^{-1}$. The dashed 
line corresponds to $\alpha_{J} - \alpha_{P}$ = (1.2 - 0.5) = -0.7, values as expected from a 
Regge model where the baryon pair production at very high energy is governed
by Pomeron exchange and baryon transport by string-junction 
exchange~\cite{kharzeev,kaidalov}. The $\alpha_{J}$ and $\alpha_{P}$ parameters 
corresponds to the junction intercept and the Pomeron intercept in the
models. The $\chi^{2}/ndf$ for the fit is 66/6. The best fit ($\chi^{2}/ndf$  = 4/5)
is however obtained for  $\alpha_{J} - \alpha_{P}$ = -0.97 $\pm$ 0.05 (solid line), 
with about a  factor of 3 increase in the value of parameter $C$ compared 
to the results presented in dashed lines. The value of $C$ for the case is
10.3 $\pm$ 0.4 and that for the second case (solid line) is 35 $\pm$ 7.
The Regge model inspired fit to the heavy-ion data (dotted line), yields a poor $\chi^{2}/ndf$  = 33/7,
with fit parameters $C$ = 119208 $\pm$ 49600 and $\alpha_{J} - \alpha_{P}$ = -2.9 $\pm$ 0.2.
The fit misses the the higher energy data points where such a model is more reliable.
Constraining the fit to the energy range of 23 to 200 GeV gives  $C$ = 205 $\pm$ 580 and 
$\alpha_{J} - \alpha_{P}$ = -1.2 $\pm$ 0.6 with $\chi^{2}/ndf$ = 0.6/1. Upcoming heavy-ion 
data at higher beam energies in LHC will give a clear picture of applicability of
such a model to heavy-ion collisions.

In summary, we have presented a compilation of the available data for
$\bar{p}/p$ ratio at midrapidity for $p$+$p$ collisions as a function 
of $\sqrt{s}$. We have also compared these ratios to the beam energy
dependence from heavy-ion collisions and found that below 
$\sqrt{s_{\mathrm {NN}}}$ = 200 GeV, the proton contribution at 
midrapidity in A+A collisions is significantly more compared to 
those in $p$+$p$ collisions.  The $\bar{p}/p$ ratio is constant as a function
of $p_{T}$ for all the beam energies for $p_{T}$ $<$ 1 GeV/$c$. This experimental
observation already puts a constrain on mechanism of baryon production such as
those implemented in HIJING/B-$\bar{\mathrm B}$. We also compared the $\bar{p}/p$ 
ratio vs. $\sqrt{s}$ to results from various models with different 
baryon production mechanisms, such as PYTHIA, PHOJET and  HIJING/B-$\bar{\mathrm B}$ 
with default settings. It is observed that  PHOJET gives the best description 
of the data for all  $\sqrt{s}$, PYTHIA gives higher values of the  
$\bar{p}/p$ ratio for $\sqrt{s}$ $<$ 200 GeV and  HIJING/B-$\bar{\mathrm B}$
under predicts the ratio for all beam energies. A detailed investigation 
of various mechanisms of baryon production as implemented in PYTHIA shows 
that the baryon production through diquark-antidiquark pair production with
the diquark being treated as a unit and the additional 
possibilities (arrived by tuning various parameters) of diquark splitting 
according to the popcorn scheme and the production of first rank baryons 
suppressed gives a reasonable description for the $\bar{p}/p$ 
ratio for $\sqrt{s}$ $<$ 200 GeV.
The asymmetry, a measure of proton stopping at the midrapidity in $p$+$p$
collisions are presented. The fraction of protons stopped around midrapidity 
varies from 46\% at $\sqrt{s}$ = 23 GeV to 0.05\% for $\sqrt{s}$ = 7 TeV. This
energy range corresponds to a range in $y_{\mathrm {beam}}$ from 3 to 9 units, 
respectively.  The data has also been compared to baryon string junction 
motivated phenomenological  function whose parameters can constrain the 
Regge-model inspired descriptions of baryon asymmetry in $p$+$p$ collisions.

\noindent{\bf Acknowledgments}\\
We thank Dr. Y. P. Viyogi for useful comments on the paper.
Financial assistance from the Department of Atomic Energy, Government 
of India is gratefully acknowledged. PKN is grateful to the Board 
of Research on Nuclear Science and Department of Atomic Energy,
Government of India for financial support in the form of Dr. K.S. Krishnan
fellowship. LK is supported by DOE grant DE-FG02-89ER40531. \\

\end{document}